\documentclass[prl,twocolumn,showpacs,amsmath,amssymb,superscriptaddress]{revtex4-1}
\usepackage{physics}
\usepackage{graphicx}
\usepackage{epsfig,epstopdf} 
\usepackage{dcolumn}
\usepackage{bm}
\usepackage{bbm}
\usepackage{ulem}
\usepackage{color}
\usepackage{slashed}
\usepackage{hyperref}
\usepackage{mathrsfs}
\usepackage{subfigure}
\usepackage{verbatim}
\usepackage{dsfont}
\usepackage{float}
\usepackage{tikz}
\usepackage{multirow}
\usepackage{etex}
\usepackage[etex=true,export]{adjustbox}

\newcommand{\beq}{\begin{equation}}
\newcommand{\eeq}{\end{equation}}
\newcommand{\beql}{\begin{equation*}}
\newcommand{\eeql}{\end{equation*}}
\newcommand{\beqn}{\begin{eqnarray}}
\newcommand{\eeqn}{\end{eqnarray}}

\renewcommand{\vec}[1]{\mbox{\boldmath$#1$}}

\begin{document}
\title{Braiding Majorana modes in spin space: from worldline to worldribbon}
\author{Xun-Jiang Luo}
\thanks{These authors contribute equally to this work.}
\affiliation{School of Physics and Wuhan National High Magnetic Field Center, Huazhong University of
Science and Technology, Wuhan, Hubei 430074, China}
\author{Ying-Ping He}
\thanks{These authors contribute equally to this work.}
\affiliation{ International Center for Quantum Materials and School of Physics, Peking University, Beijing 100871, China}
\affiliation{Collaborative Innovation Center of Quantum Matter, Beijing 100871, China}
\author{Ting Fung Jeffrey Poon}
\affiliation{ International Center for Quantum Materials and School of Physics, Peking University, Beijing 100871, China}
\affiliation{Collaborative Innovation Center of Quantum Matter, Beijing 100871, China}
\author{Xin Liu}
\email{phyliuxin@hust.edu.cn}
\affiliation{School of Physics and Wuhan National High Magnetic Field Center, Huazhong University of
Science and Technology, Wuhan, Hubei 430074, China}
\author{Xiong-Jun Liu}
\email{xiongjunliu@pku.edu.cn}
\affiliation{ International Center for Quantum Materials and School of Physics, Peking University, Beijing 100871, China}
\affiliation{Collaborative Innovation Center of Quantum Matter, Beijing 100871, China}
\date{\today}

\begin{abstract}
We propose a scheme to braid Majorana zero modes (MZMs) by steering the spin degree of freedom, without moving, measuring, or more generically fusing the modes. For a spinful Majorana system, we show that braiding two MZMs is achieved by locally winding the Majorana spins, which topologically corresponds to twisting two associated worldribbons, equivalent to worldlines that track the braiding history of MZMs.
We demonstrate the feasibility of applying the current scheme to the superconductor/2D-topological-insulator/ferromagnetic-insulator (SC/2DTI/FI) hybrid system which is currently under construction in experiment.
The single (or full) braiding of two MZMs is precisely achieved by adiabatically winding the FI magnetization by a half (or complete) circle, with the braiding operation shown to be robust against local imperfections such as irregular winding paths, the static and dynamical disorder effects. The stability is a consequence of the intrinsic connection of the current scheme to topological charge pumping. The proposed device involves no auxiliary MZMs, rendering a minimal scheme for observing non-Abelian braiding and having advantages with minimized errors for the experimental demonstration.
\end{abstract}
\maketitle

\textit{Introduction -}The most exotic property of Majorana zero mdoes (MZMs) is embedded in its non-Abelian braiding statistics \cite{Kitaev2001,Ivanov2001,Nayak2008}, which is important for fundamental physics and also has potential application to topological quantum computation (TQC). The remarkable progresses in the recent experiments \cite{Mourik2012,Deng2012,Rokhinson2012,Das2012,Wang2012,Churchill2013,Xu2014,Nadj-Perge2014,Chang2015,Albrecht2016,Wiedenmann2016,Bocquillon2016,Zhang2017} of observing MZMs bring us closer to detecting their non-Abelian statistics, which is a smoking gun for their existence. The most straightforward way of braiding two anyons is to physically move one around the other in real space. Various superconducting junctions such as T-junction \cite{Alicea2011,Liu2014}, Y-junction \cite{Sau2011,Heck2012,Liu2013PhysRevB,Hyart2013,Wu2014}, $\pi$-junction \cite{Heck2015} and U-junction \cite{Liu2016,Karzig2017} are proposed to move MZMs by coupling them in certain order through tuning a series of gates. Recently, it is also shown that braiding MZMs can be realized through measuring their fusion results and keeping the desired data \cite{Vijay2016}. All these methods can be classified as fusion-based braiding, since they rely on fusing (or equivalently coupling) different MZMs, which (effectively) transports MZMs under a controllable way. Note that the transporting or fusion operations typically cause complexity in the manipulation across junctions or uncontrollable errors during the fusion-measurement processes, which brings challenges for the experimental identification of non-Abelian statistics. On the other hand, from TQC theory we know that if anyons have internal degree of freedom, e.g., the flux-charge composite model \cite{Preskill2004}, the associated worldlines, which characterize the braiding trajectories of anyons, can be extended to worldribbons which are called framing \cite{Nayak2008}. Braiding two worldribbons, corresponding to exchanging two anyons with a given fusion channel, is equivalent to twist locally each worldribbon around itself~
\cite{Preskill2004,Finkelstein1968,topspin,Pachos2012}. This suggests fusion-free schemes to braid anyons, as applied to the Majorana system proposed in the present study.

In this work, we propose to braid MZMs in solid state systems by adopting manipulation on the spin degree of freedom of Majorana modes. With two generic theorems shown here, we demonstrate that the single (or full) braiding of two MZMs can be achieved by adiabatically winding their spins by a half (or full) circle. The braiding operation is topologically related to twisting two associated worldribbons, equivalent to worldlines which track the braiding history of MZMs. The application of the current scheme to the superconductor/2D-topological-insulator/ferromagnetic-insulator (SC/2DTI/FI) hybrid system is proposed and studied in detail. Without the need of moving or measuring the MZMs, the explicit advantages of the present fusion-free scheme are revealed with analytical and numerical results.
%

\begin{figure}
\includegraphics[width=0.9\columnwidth]{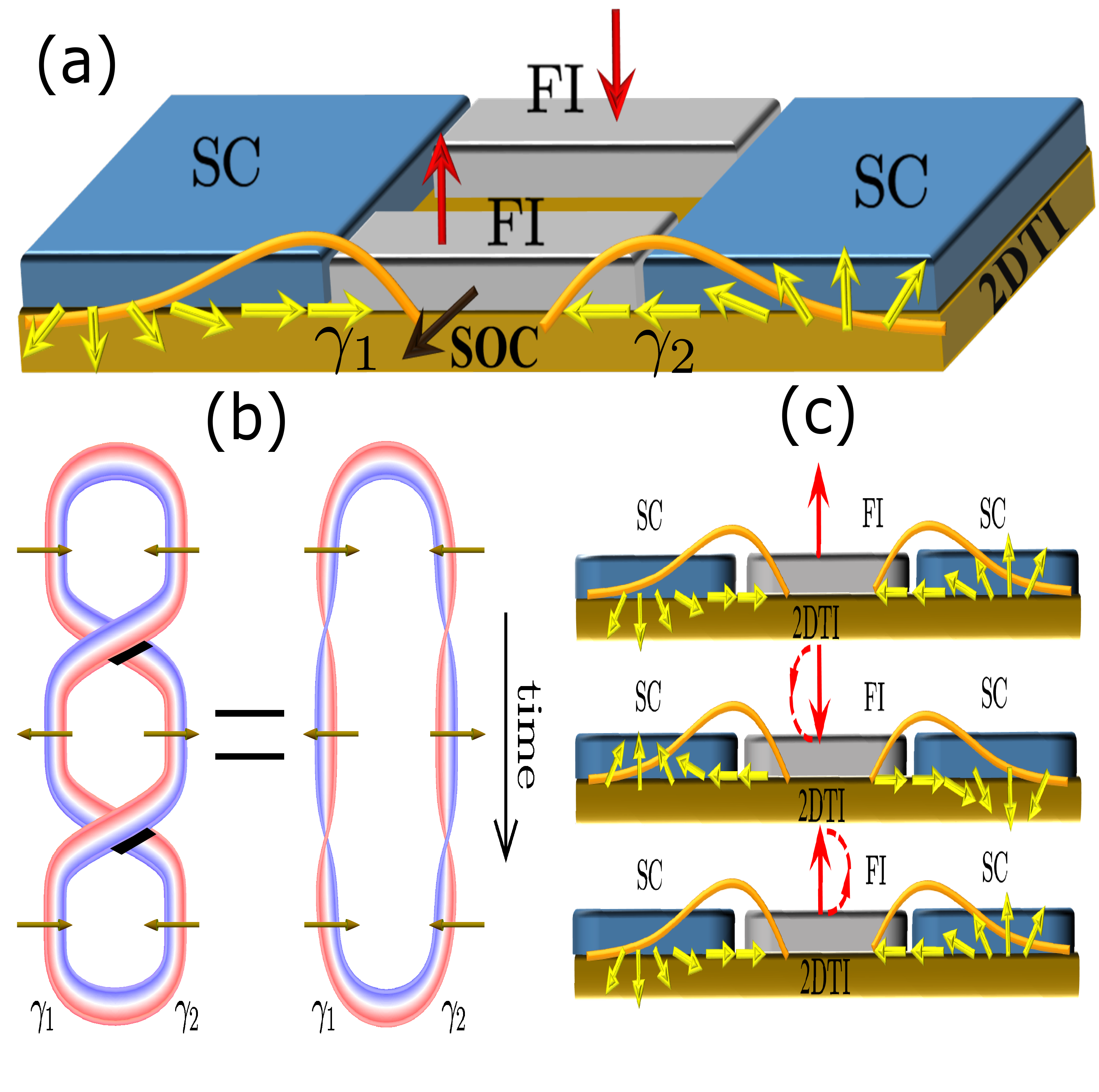}
\caption{ {(a)} The MZMs at the interface of the SC/FI/SC interfaces on the top of a QSH system generally have spatially dependent spin polarization. {(b)} The monodromy operator can be realized by either braiding two MZMs or twist each worldribbons by $2\pi$. The arrows indicate the MZM spin. The blue and red edges of the ribbon denote the evolution of internal degree of freedom. (c) The spin texture of the Majorana spins during rotating FI magnetization by $2\pi$.}
\label{ribbon}
\end{figure}

\textit{Braiding MZMs in spin space-}
We start with a quasi-1D topological superconductor (TSC), realized via nanowires or edges of a 2D TI, with the Hamiltonian in spinor basis $\hat{c}(\bm{r})=\bigr[c_{\uparrow}(\bm{r}),c_{\downarrow}(\bm{r}),c^{\dagger}_{\downarrow}(\bm{r}),-c^{\dagger}_{\uparrow}(\bm{r})\bigr]^T$ given by
\beqn
\label{ham-1}
\hat{H}=\left(\begin{array}{cc}
		h(\bm{\hat{p}})+\bm{m}(\bm{r})\cdot\sigma&\Delta_{\rm SC}(\bm{r})\\
		\Delta^{\dagger}_{\rm SC}(\bm{r}) &-h(\bm{\hat{p}})+\bm{m}(\bm{r})\cdot\sigma
	\end{array}\right),
\eeqn
where $\sigma_{x,y,z}$ are Pauli matrices in spin space, $h(\bm{\hat{p}})$ is time-reversal invariant and its explicit form for concrete example will be given later, the proximity induced $s$-wave SC order $\Delta(\bm{r})$ and Zeeman term $\bm{m}(\bm{r})$ are spatial dependent and determine topological/trivial regions. A MZM exists at an interface between such two regions. The particle-hole symmetry enforces the electron and hole components of a MZM to have identical spin polarization \cite{He2014,Liu2015}, with the Majorana wave-function~\cite{Halperin2012}
\beqn\label{SH-1}
\Psi({\bm{r}})=(\psi_{\rm e}(\bm{r}), i\sigma_y \psi_{\rm e}^{*}(\bm{r})^{\rm T},
\eeqn
where the two-component spinor $\psi_{\rm e}$ determines the spatial distribution of Majorana spin polarization. Before moving to the discussion on specific system, we show first the generic results of braiding two MZMs $\gamma_1$ and $\gamma_2$, separated by a magnetization ($\bm m$) dominated region, by steering the magnetization between them.

\textbf{Theorem 1:} {\it The adiabatic spin evolution of each MZM, following an arbitrary closed path in varying the direction of $\bm m$ without closing bulk gap, accumulates a geometric phase quantized to $n\pi$, which leads to $n$ times full braiding of $\gamma_1$ and $\gamma_2$ in fusion space.}

\textbf{Theorem 2:} {\it The adiabatic evolution of MZMs $\gamma_1$ and $\gamma_2$ following an arbitrary magnetization winding path, with the initial and final Zeeman term satisfying $\bm{m}_{\rm i}=-\bm{m}_{\rm f}$, reverses the spin of each MZM, which corresponds to a single braiding of $\gamma_1$ and $\gamma_2$.}


The two theorems are generic, independent model details, while we consider the SC/2DTI/FI hybrid system for convenience [Fig.~\ref{ribbon}(a)]. For theorem I, we consider Majorana evolution by tuning the direction of magnetization at the bottom edge [Fig.~\ref{ribbon}(a)] along an arbitrary closed trajectory from time $t=0$ to $t=T_0$. The accumulated phase for the closed evolution trajectory consist of dynamics phase, Berry phase and monodromy phase. The dynamic phase vanishes due to the zero eigenenergy of MZMs. The Berry connection for the instantaneous MZM eigen-function given in Eq.~\eqref{SH-1} also vanishes because
$\rm{Im}\bra{\Psi}\partial_t \ket{\Psi}=\rm{Im} \left( \bra{\psi_e}\partial_t \ket{\psi_e} + \bra{\psi_e^*} \partial_t \ket{\psi_e^*} \right)=0$.
Thus the accumulated phase is completely contributed from the monodromy phase, say the evolution of $\psi_e$ in the Majorana spin space. This follows that $\langle\Psi(T_0)|\Psi(0)\rangle=\pm 1$, showing that the solid angle enclosed by the Majorana spin trajectory generically takes $2n\pi$, corresponding to monodromy phase $n\pi$, even though the solid angle enclosed by the magnetization trajectory can be arbitrary. The Majorana spin evolution implies that the world lines, tracking the trajectories of MZM evolution in spacetime, should be extended to world ribbons [Fig.~\ref{ribbon}(b)] \cite{Preskill2004,Finkelstein1968,topspin,Pachos2012} with appropriate framing \cite{Nayak2008} in spin space. For $n=1$, we have $\gamma_{1,2}(T)= -\gamma_{1,2}(0)$, giving the full braiding operation $\exp(-\pi \gamma_1 \gamma_2 /2)$ \cite{Alicea2012}.
According to the spin-statistics theorem \cite{Finkelstein1968,topspin}, twisting each world ribbon of two MZMs by $2\pi$ is identical to a full braiding [Fig.~\ref{ribbon}(b) and ~\ref{ribbon}(c)], providing the unambiguous framing choice. Generically, the $2n\pi$ rotation of MZM spin corresponds to $n$ times full braiding. This proves theorem 1.

\begin{figure*}
\centering
\includegraphics[width=1.7\columnwidth]{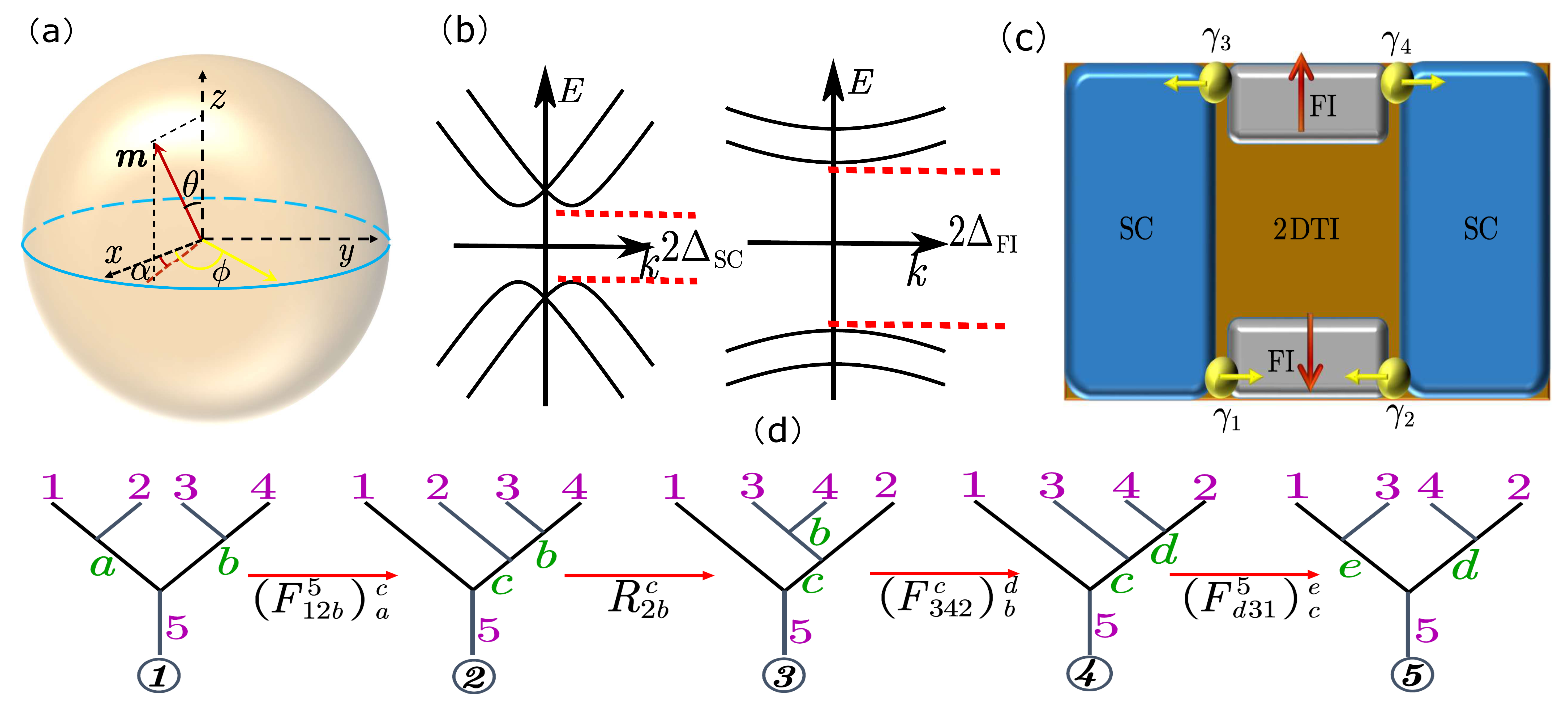}
\caption{{(a)} The relations among the Majorana spin in FI region (yellow arrow), magnetization (red arrow) and SOC field direction ($\bm{e}_z$). The polar and azimuth angles of FI magnetization are $\theta$ and $\alpha$, respectively. The Majorana spin is perpendicular to the SOC field direction with azimuth angle $\phi$. (b) The dispersion of the electron and hole under SC and FI. The left and right plots are the band structure of the edge states underneath the superconductor and ferromagnetic insulator with band gap $\Delta_{\rm SC}$ and $\Delta_{FI}=(|\bm{m}|\sin\theta-\mu)$. (c) Majorana qubits in the SC/QSH/FI hybrid system. The yellow (red) arrows represent the directions of local spin polarizations for MZMs (FI magnetization). For simplicity, it is taken that $m_{z}=0$ and $\mu=0$. {(d)}The transformation of the two fusion spaces through four $F$ matrices and one $R$ matrix.  }
\label{MZM-spin}
\end{figure*}

We show further theorem II from the MZM evolution $\Psi(T_0/2)=\hat{U}(T_0/2) \Psi(0)$ by tuning
the magnetization to $\bm{m}(T_0/2)=-\bm{m}(0)$, with $\hat{U}(T_0/2)$ the MZM unitary evolution matrix from $t=0$ to $t=T_0/2$. As only the Zeeman term in the Hamiltonian~\eqref{ham-1} breaks time-reversal symmetry, we have
\beql
\hat{H}(0)\Psi(0)=\hat{T}\hat{H}(0)\Psi(0)=\hat{H}(T_0/2)\hat{T}\Psi(0)=0,
\eeql
for which $\hat{T}\Psi(0)$ is the MZM at time $T_0/2$ and the Majorana spin at $t=T_0/2$ is opposite to its initial direction [Fig.~\ref{ribbon}(c)]~\cite{Supp}. Moreover, as $\Psi(t)$ and $\hat{T}\Psi(t)$ are MZMs for the TSC Hamiltonian in Eq.~\eqref{ham-1} with opposite magnetization, the evolution of $\Psi(t)$ and $\hat{T}\Psi(t)$ have the same unitary evolution matrix~\cite{Supp} so that
 \beqn\label{Ta}
 \hat{T}\Psi(T_0/2)=\hat{U}(T_0/2) \hat{T}\Psi(0).
 \eeqn
 It is noted that as long as $\Psi(0)$ takes a Majorana form, both $\Psi(T_0/2)$ and $\hat{T}\Psi(0)$ are also Majoranas, hence
 \beqn\label{Tb}
 \hat{T}\Psi(0)=\zeta\Psi(T_0/2), \ \ \zeta=\pm 1.
 \eeqn
 Combining Eq.~\ref{Ta} and Eq.~\ref{Tb}, we have \cite{Supp}
 \beqn
 \hat{U}^2(T_0/2) \Psi(0)=\hat{T^2}\Psi(0)= -\Psi(0).
 \eeqn
Thus the adiabatic evolution matrix $\hat{U}^2(T_0/2)$ is equivalent to odd times full braiding according to theorem 1. Without loss of generality, we consider a single full braiding. In the Fermion parity basis $-i\gamma_1\gamma_2=\pm$, the evolution operator $\hat{U}(T_0/2)$ is a diagonal
matrix with $\hat{U}^2(T_0/2)=\exp(-i s_z \pi/2)$, which is followed that $\hat{U}(T_0/2)=\exp(-i s_z \pi/4)=\exp(-\pi \gamma_1\gamma_2/4)$ with $s_z$ the Pauli matrix in fusion space. This is the single
braiding operator, equivalent to the ribbon equation $\exp(-i \pi S_{\gamma_1}-i\pi S_{\gamma_2} + i\pi S_{g})$ with $S_{\gamma_{1,2};g}$ the topological spins of MZMs and the fusion outcome respectively, which corresponds to twisting the world ribbon by $\pi$ and is consistent with the framing choice. Thus adiabatically reversing the magnetization $\bm{m}(T_0/2)=-\bm{m}(0)$ gives odd times single braiding operation, completing the proof of theorem 2.

It is instructive to apply the above generic theorems to a concrete 1D model of the SC/2DTI/FI hybrid system and show the Majorana braiding by spin manipulation (Fig.~\ref{ribbon}(a)). Around the Fermi energy which is inside the TI bulk gap, the single-particle Hamiltonian reduces to $h(\hat{p})=v_{\rm f} \hat{p}\sigma_{z}-\mu$, where $v_{\rm f}$ is the Fermi velocity of edge states and $\mu$ is the chemical potential. Let the magnetization have a polar angle $\theta$, which gives the in-plane component $m_{\parallel}=|\bm{m}|\sin\theta$ (Fig.~\ref{MZM-spin}(a)). In the case of $m_{\parallel}^2 > \mu^2+\Delta_{\rm SC}^2$ \cite{Jiang2013} (Fig~\ref{MZM-spin}(b)), at each SC/FI interface locates a single MZM, e.g., the MZMs $\gamma_1$ and $\gamma_2$, as shown in Fig.~\ref{MZM-spin}(c). The Majorana spin is polarized perpendicular to the spin-orbit ($z$) axis. In the FI region, the electron part of MZM reads (Fig.~\ref{MZM-spin}(c))~\cite{Supp}
\begin{eqnarray*}
\psi_{e}^{1,2}= \frac{e^{i\frac{\pi}{4}}}{\sqrt{2}} \left( \begin{array}{c} e^{-i\phi^{1,2}/2} \\ e^{i\phi^{1,2}/2} \end{array} \right),  \ \phi^{1,2}= \alpha\pm\cos^{-1}\left(\frac{ \mu}{m_{\parallel}} \right),
\end{eqnarray*}
with $\phi^{1,2}$ and $\alpha$ the azimuthal angles of the Majorana spins and the magnetization, respectively (Fig.~\ref{MZM-spin}(a)). In SC region, the Majorana spin forms a helical texture in the $x-y$ plane (Fig.~\ref{ribbon}(a)). A key observation is that when the direction of magnetization $\bm m$ varies by one closed trajectory, as long as the trajectory encloses the spin-orbit ($z$) axis, the Majorana spin winds in the $x-y$ plane and spans $2\pi$ solid angle in the Bloch sphere (Fig.~\ref{MZM-spin}(a)), yielding $\pi$ geometric phase as stated in
theorem 1. On the other hand, tuning FI magnetization from $\bm{m}(0)$ to $\bm{m}(T_0/2)=-\bm{m}(0)$ leads to $\phi^{1,2}(T_0/2)-\phi^{1,2}(0)=\pi$, for which the Majorana spin reverses sign, consistent with the theorem 2.

{\it Topological pumping-} The physics behind the equivalence between braiding MZMs and rotating MZM spins is more transparent when adopting another two MZMs $\gamma_{3,4}$ (Fig.~\ref{MZM-spin}(c)). The four MZMs, with fixed total fermion parity, can form one qubit which are germinated from two complex fermion modes, with the fermion operators and fusion states being shown in Tab.~\ref{table:table1}. The nonlocal fermion operators $f_{\rm u}$ and $f_{\rm d}$ ($d_{\rm L}$ and $d_{\rm R}$) are constructed from the two MZMs attached to the upper and lower FI islands (left and right SC islands), respectively.
As our system has only four MZMs which can be well separated, adiabatically rotating the magnetization at the bottom edge (Fig.~\ref{MZM-spin}(c)) does not change the local fermion parity defined by $i\gamma_3\gamma_4$, nor the fermion parity defined by $i \gamma_1 \gamma_2$. Thus in FI basis, according to theorem 2, adiabatically reverse the magnetization at the bottom once and twice correspond to the evolution of the qubit state with an diagonal matrix $\exp(-is_z \pi/4)$ and $\exp(-i s_z \pi/2)$ respectively. On the other hand,
the FI and SC basis are related through $F$ and $R$ matrices \cite{Preskill2004,Kitaev2006} (Fig.~\ref{MZM-spin}(d)) which result in the transformation
\beqn\label{trans}
\left( \begin{array}{c} |00\rangle_{\rm FI} \\ |11\rangle_{\rm FI}  \end{array}\right)= \hat{T} \left( \begin{array}{c} |00\rangle_{\rm SC} \\ |11\rangle_{\rm SC}  \end{array}\right), \ \ \hat{T}=\frac{1}{\sqrt{2}}\left( \begin{array}{cc} 1 & 1 \\ i & -i   \end{array}\right).
\eeqn
Accordingly, the single braiding and full braiding matrices in SC basis take
\beql
\exp(-\frac{\pi \gamma_1 \gamma_2}{4})=e^{-i\frac{\pi}{4}s_x}, \ \ \exp(-\frac{\pi \gamma_1 \gamma_2}{2})=e^{-i\frac{\pi}{2}s_x}.
\eeql
If the initial Majorana qubit is in $|00 \rangle_{\rm SC}$, adiabatically rotating the FI magnetization by $2\pi$ makes the final state evolve to $|11\rangle_{\rm SC}$ which corresponds to the fermion parity switch between the left and right hand SCs (Fig.~\ref{MZM-spin}(c)). This is a consequence of the Thouless charge pumping~\cite{Thouless1983,Qi2008,Xiao2010,Avron2004}, which pumps a single electron between the left and right hand SCs through rotating magnetization angle by $2\pi$ in the FI region, accounting for the robustness of the full braiding operation in the present proposal. Further, adiabatically rotating the FI magnetization by $\pi$ is equivalent to braiding $\gamma_1$ and $\gamma_2$ once which leads to the final state $(\ket{00}_{\rm SC}-i\ket{11}_{\rm SC})/\sqrt{2}$. The fermion parity in the left and right SCs has $50\%$ probability to be switched, which corresponds to the half charge pumping through flipping magnetization~\cite{Supp}. The connection of braiding to the quantized charge pumping can be observed by measuring the fermion parity in the two SC islands in the Coulomb blockade regime~\cite{Aasen2016,Karzig2017}. Note that tuning magnetization to $\bm{m}(T_0/2)=-\bm{m}(0)$ can be achieved with standard technique by setting $\bm{m}(0)$ along the easy axis of the FI and switching its direction.

\begin{figure}
\centering
\includegraphics[width=3.2in]{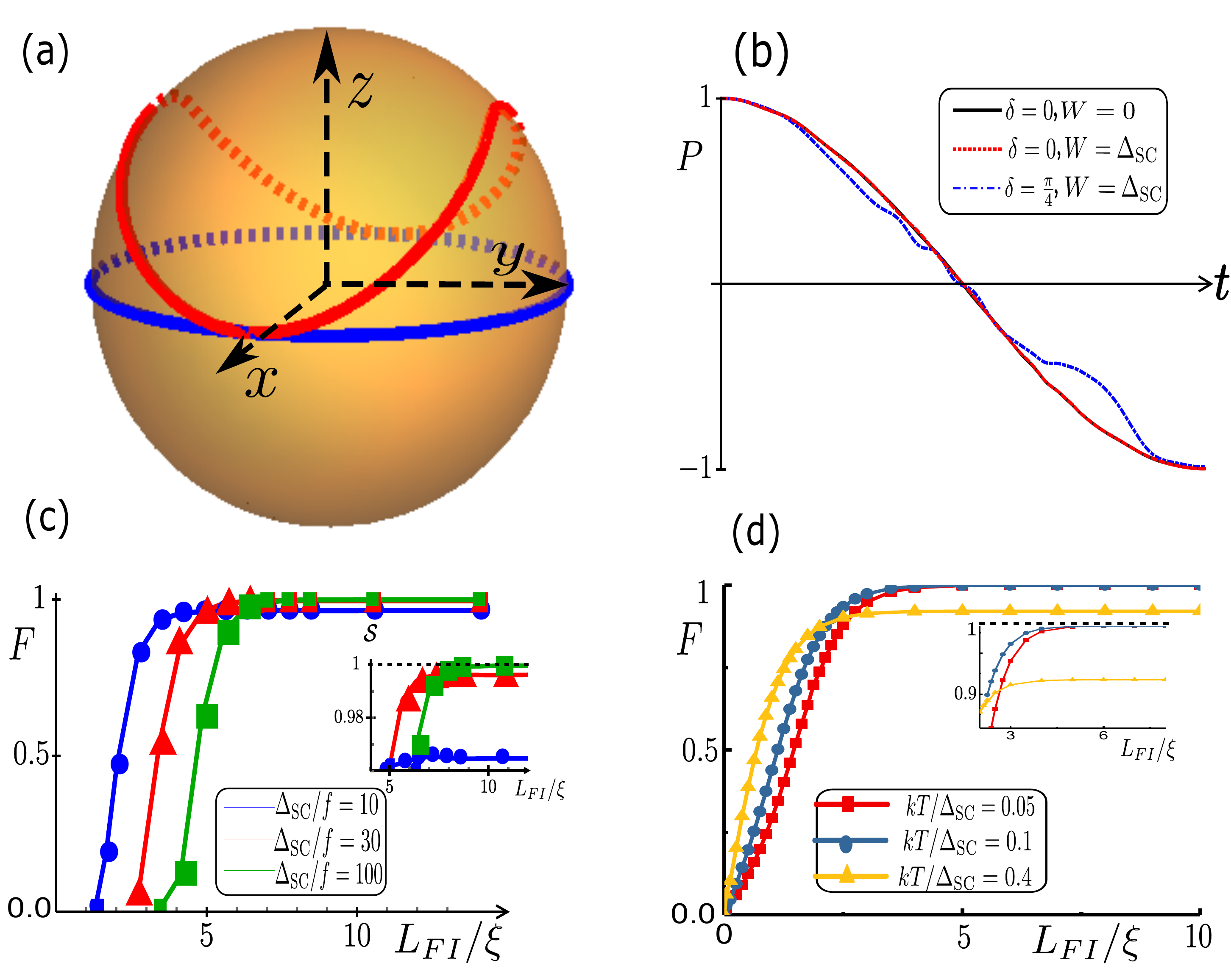}
\caption{(a) The two magnetization trajectories in the numerical simulation. (b) The fermion parity polarization for various magnetization evolution paths and impurity strengths. (c) The fidelity of the full braiding operation versus FI length for various braiding time with $f=\omega/2\pi$. (d) The fidelity of the full braiding operation versus FI length with different temperatures.}
\label{braiding}
\end{figure}

\begin{table}
  \begin{center}
  \begin{adjustbox}{max width=0.5\textwidth}
  \begin{tabular}{|c|c|l|}
  \hline
 basis & fermion operators & fusion states \\
  \hline
 $|i \gamma_1 \gamma_2,$ & $f_{\rm u}=(\gamma_3 + i\gamma_4)/2$, & $ |00\rangle_{\rm FI} ,$\\
 $i\gamma_3 \gamma_4 \rangle$ &$f_{\rm d} = (\gamma_1 + i\gamma_2)/2$ & $ |11\rangle_{\rm FI}=f^{\dagger}_{\rm d}f^{\dagger}_{\rm u}|00\rangle_{\rm FI} $ \\
 \hline
 $|i\gamma_1 \gamma_3,$ & $d_{\rm L}=(\gamma_1 + i\gamma_3)/2$ & $ |00\rangle_{\rm SC} $\\
$ i\gamma_4 \gamma_2 \rangle$ &$d_{\rm R} = (\gamma_4 + i\gamma_2)/2$  & $ |11\rangle_{\rm SC}=d^{\dagger}_{\rm L}d^{\dagger}_{\rm R}|00\rangle_{\rm SC} $\\
 \hline
\end{tabular}
\end{adjustbox}
\caption{Two different fermion parity basis. The fermion modes $f_u$ and $f_d$ are defined corresponding to the FI, while $d_L$ and $d_R$ are defined corresponding to the SC.}
\label{table:table1}
\end{center}
\end{table}

{\it Numerical simulations-} Now we consider the numerical simulation of the present braiding scheme by taking $h(\hat{\bm{p}})$ as the Bernevig-Hudges-Zhang Hamiltonian for 2D TI~\cite{Bernevig2006,Supp}. For generality, we add to the system spin independent disorders with random disorder strength in the range $[-W,W]$, and tune the magnetization as
\beql
\vec{m}(t)=|\bm{m}|\bigr[\cos\delta(t) \cos(\omega t) ,  \cos\delta(t) \sin(\omega t),  \sin\delta(t)\bigr],
\eeql
where $\omega=2\pi/T_0$, and $\delta(t)=0$ ($\frac{\pi}{4}\sin^2(\omega t)$) corresponds to a regular (irregular) magnetization tuning trajectory [Fig.~\ref{braiding}(a)]. The fidelity of braiding operation is quantified by the MZM wave function overlap $P=\langle\Psi_1(t)|\Psi_1(0)\rangle$  (e.g. for $\gamma_1$), which is real and plotted in Fig.~\ref{braiding}(b). At $t=T_0/2$ ($t=T_0)$, all the curves converge to $P=0$ ($P=-1$), showing that MZM spin is reversed (acquires $\pi$ geometry phase), which gives the single braiding (full braiding) operation. Importantly, the numerical results show that braiding is robust against disorder effects, and is not affected by varying the magnetization  trajectories. The results have been further confirmed by considering dynamical noise in the magnetization trajectories (see Supplementary Materials~\cite{Supp}). The braiding error may be caused by non-adiabatic manipulation and thermal effects, as shown in in Fig.~\ref{braiding}(c) and Fig.~\ref{braiding}(d), respectively, where we calculate the Fermion parity switch $\delta F$ after a full braiding~\cite{Supp}. Interestingly, the deviation of Fermion parity switch from unity is dramatically suppressed in both cases through increasing the FI length. Moreover, the thermal excitations in the FI region are suppressed by Zeeman gap in the FI region $\Delta_{\rm FI}=(|\bm{m}| \sin\theta -\mu)$, which is typically larger than the SC gap and can improve the validity of adiabatic condition.

Before conclusion we discuss the experimental setup for the realization. The 2DTI has been realized in HgTe/CdTe \cite{Konig2007,Bocquillon2016,Deacon2017} and InAs/GaSb \cite{Knez2011,Du2015,Pribiag2015} heterostructures. The $4\pi$-periodicity Josephson effect has been observed in superconducting proximity coupled HgTe/CdTe quantum well \cite{Wiedenmann2016}. Recently, the 2DTI, superconductivity, and FI are observed in single-layer van der Waals crystals such as WeTe2 \cite{Fei2017,Peng2017}, NbSe$_2$ \cite{Xi2015} and CrI$_3$ \cite{Huang2017} respectively, which exhibit great advantages in fabricating FI-SC junction on 2DTI surface due to the vdW stacking.
The relevant experimental parameters of typical materials are estimated as follows. The ferromagnetic insulator, such as YIG~\cite{Singh2017}, can induce an effective exchange field up to $1T$ into 2D material, which corresponds to a spin splitting gap $|\bm{m}|=3$meV~\cite{Hu2016} of the 2DTI edge state when the magnetization is perpendicular to the SOC field, greater than the typical proximity induced SC gap $\Delta=0.1$meV. For the Fermi velocity $\hbar v_{\rm f}=0.36\textrm{meV} \cdot\mu\textrm{m} $\cite{Nilsson2008}, the FI coherence length is about 0.12$\mu$m, which implies that the braiding can be well achieved with negligible error when the FI length is over 1$\mu$m, according to the simulation in Fig.~\ref{braiding}.

{\it Conclusion-} We have proposed a new scheme to braid MZMs by steering spin degree of freedom of Majoranas, different from the conventional schemes which rely on moving, measuring, or more generically fusing the MZMs. We applied the new scheme to the SC/2DTI/FI hybrid system, and demonstrated with experimental feasibility the non-Abelian braiding of MZMs by locally winding FI magnetization. The proposed device involves no auxiliary MZMs, rendering a minimal scheme of observing non-Abelian statistics and having advantages with minimized errors in experimental demonstration, and shall open up fusion-free approaches within current experimental accessibility to probe MZM braiding statistics.

We would like to thank Ruirui Du, Hong Ding, Dong-Ling Deng, Meng Cheng, Jainendra Jain, Zheng-Xin Liu, Chao-Xing Liu, Xiaopeng Li, Zhenhua Qiao and Kunhua Zhang for useful discussions. This work is supported by National Key R\&D Program of China (Grant No. 2016YFA0401003 and 2016YFA0301604),  NSFC (Grant No.11674114, No. 11574008, and No. 11761161003), and Thousand-Young-Talent program of China.

%

\pagebreak
\pagebreak
\clearpage
\setcounter{equation}{0}
\setcounter{figure}{0}

\renewcommand{\theparagraph}{\bf}
\renewcommand{\thefigure}{S\arabic{figure}}
\renewcommand{\theequation}{S\arabic{equation}}
\onecolumngrid
\flushbottom

\section{SUPPLEMENTARY MATERIAL}

\section{Proof of theorem 2}

The Hamiltonian for a generic spinful topological superconductor in the spinor basis $\hat{c}(\bm{r})=\left(c_{\uparrow}(\bm{r}),c_{\downarrow}(\bm{r}),c^{\dagger}_{\downarrow}(\bm{r}),-c^{\dagger}_{\uparrow}(\bm{r})\right)$ is
\begin{eqnarray}
\label{Ham-1}
\hat{H}=\left(\begin{array}{cc}
		h(\hat{p})-\mu+\bm{m}(\bm{r})\cdot\sigma&\Delta_{\rm SC}(\bm{r})\\
		\Delta_{\rm SC}(\bm{r}) &-h(\hat{p})+\mu+\bm{m}(\bm{r})\cdot\sigma
	\end{array}\right),
\end{eqnarray}
where $\sigma_{x,y,z}$ are Pauli matrices in spin space, $h(\hat{p})$ is a generic time-reversal invariant Hamiltonian, $\mu$ is chemical potential, $\Delta_{\rm SC}(\bm{r})$ and $\bm{m}(\bm{r})$ are the superconductor gap and magnetization, respectively. We denote the  $i$th MZM wave function as $\Psi_i(\bm{r},t)$ and consider that all MZMs are isolated from each other so that they have exactly zero energy. When we adiabatically rotate the magnetization $\bm{m}$, the MZM wave function evolves as
\beqn\label{evo-1}
\Psi_i(\bm{r},t)=\mathcal{T} \exp\left(-i\int_{0}^t \frac{\hat{H}(\bm{m}(t\rq{}))}{\hbar} dt\rq{}\right)\Psi_{i}(\bm{r},0)=U(t,0) \Psi_{i}(\bm{r},0),
\eeqn
where $\mathcal{T}$ denotes time order. Taking an infinitesimal evolution time $t=\delta t$, the Majorana wave function up to the first order is given by
\beqn\label{evo-1}
\Psi_i (\bm{r},\delta t) &=& U(0,\delta t) \Psi_i (\bm{r},0) \approx \left(1-\frac{i}{\hbar}\int_{0}^{\delta t} \hat{H}(\bm{m}(t)) dt \right) \Psi_i(\bm{r},0) \nonumber \\
&=& \left(1-\frac{i}{\hbar}\int_{0}^{\delta t} \big[\hat{H}(\bm{m}(0)) + \bm{m}(t)\cdot \bm{\sigma}-\bm{m}(0) \cdot \bm{\sigma} \big] dt \right)  \Psi_i(\bm{r},0) \nonumber \\
& = & \left(1-\frac{i}{\hbar}\int_{0}^{\delta t} \big[ \bm{m}(t)\cdot \bm{\sigma}-\bm{m}(0) \cdot \bm{\sigma} \big] dt \right)  \Psi_i(\bm{r},0).
\eeqn
For the last equals sign, we use the fact that $\hat{H}(\bm{m}(0))\Psi_i(\bm{r},0)=0$.
In Majorana form, the $i$th instantaneous zero modes of the system satisfies
\beqn\label{ini-1}
\Psi_{i}(\bm{r},t)=\left(\begin{array}{c} \psi_{i,e}(\bm{r},t) \\ \psi_{i,h}(\bm{r},t) \end{array} \right)=\left(\begin{array}{c} \psi_{i,e}(\bm{r},t) \\ i\sigma_y \psi_{i,e}^{*}(\bm{r},t) \end{array} \right)=\left(\begin{array}{c} \psi_{i,e}(\bm{r},t) \\ \hat{T} \psi(\bm{r},t) \end{array} \right), \ \ \psi_{i,h}(\bm{r},t)=\hat{T} \psi_{i,e}(\bm{r},t), \ \ \hat{T}=i\sigma_y K.
\eeqn

The MZM wave function at $t=0$ satisfies the equation
\beqn\label{MZM-1}
\left(h(\hat{p})-\mu+\bm{m}(0)\cdot \bm{\sigma}\right) \Psi(\bm{r},0) + \Delta \hat{T} \Psi(\bm{r},0)=0.
\eeqn
If multiplying the time-reversal operator on both sides of the above equation, we have
\beqn\label{MZM-2}
&&\hat{T}\left(h(\hat{p})-\mu+\bm{m}(0)\cdot \bm{\sigma}\right) \Psi(\bm{r},0)\hat{T}^{-1} \hat{T} \Psi(\bm{r},0)+\Delta \hat{T} (\hat{T}\Psi(\bm{r},0))\nonumber \\
&&=\left(h(\hat{p})-\mu-\bm{m}(0)\cdot \bm{\sigma}\right) \Psi(\bm{r},0) (\hat{T} \Psi(\bm{r},0))+\Delta \hat{T} (\hat{T}\Psi(\bm{r},0))=0,
\eeqn
which implies that $\hat{T}\Psi(\bm{r},,0)$ is the MZM wave function when the magnetization is rotated from its initial direction to its opposite direction. Besides, it is easy to check that the electron and hole components of wave function $\hat{T}\Psi(\bm{r},,0)$ also take Majorana form.
The Majorana wave function $\Psi(\bm{r},t)$ satisfies
\beql
i\hbar \partial_t \Psi(\bm{r},t) = \hat{H}(\bm{m}(t)) \Psi(\bm{r},t).
\eeql
Multiplying $\hat{T}$ on both sides, we have 
\beql
\hat{T} i\hbar \partial_t \Psi(\bm{r},t) = -i \hbar\partial_t \hat{T}\Psi(\bm{r},t)= \hat{T} \hat{H}(\bm{m}(t)) \hat{T}^{-1} \hat{T} \Psi(\bm{r},t)=\hat{H}(-\bm{m}(t))\hat{T} \Psi(\bm{r},t).
\eeql
Thus the wave function, $\Phi(\bm{r},t)=\hat{T}\Psi(\bm{r},t)$ satisfies the equation
\beql
i\hbar \partial_t \Phi(\bm{r},t)=-\hat{H}(-\bm{m}(t))\Phi(\bm{r},t), \ \ \Phi(\bm{r},t)= \mathcal{T} \exp(i \int_0^t \hat{H}(-\bm{m}(t)) dt) \Phi(\bm{r},0).
\eeql
Forthe infinitesimal evolution time $t=\delta t$, we have
\beqn\label{S-P}
\Phi(\bm{r},\delta t) &\approx& \left(1+i \int_0^{\delta t} \hat{H}(-\bm{m}(t))dt \right) \Phi(\bm{r},0)= \left(1+\frac{i}{\hbar}\int_{0}^{\delta t} \big[\hat{H}(-\bm{m}(0)) - \bm{m}(t)\cdot \bm{\sigma}+\bm{m}(0) \cdot \bm{\sigma} \big] dt \right)  \Phi(\bm{r},0) \nonumber \\
&=& \left(1+\frac{i}{\hbar}\int_{0}^{\delta t} \big[ -\bm{m}(t)\cdot \bm{\sigma}+\bm{m}(0) \cdot \bm{\sigma} \big] dt \right) \Phi(\bm{r},0)= \left(1-\frac{i}{\hbar}\int_{0}^{\delta t} \big[\bm{m}(t)\cdot \bm{\sigma}-\bm{m}(0) \cdot \bm{\sigma} \big] dt \right) \Phi(\bm{r},0).
\eeqn
In the above derivative, we have applied the fact that $\hat{H}(-\bm{m}(0))\Phi(\bm{r},0)=0$. Comparing Eq.~\eqref{evo-1} with Eq.~\eqref{S-P}, we conclude that for MZMs, the evolution matrices of $\Psi(\bm{r},t)$ and $\Phi(\bm{r},t)=\hat{T}\Psi(\bm{r},t)$ are identical. Taking the evolution from $t=0$ to $t=T_0/2$ for example, we have 
\beqn\label{S0}
\Psi(\bm{r},T_0/2)=\hat{U}(T_0/2) \Psi(\bm{r},0), \ \  \hat{T}\Psi(\bm{r},T_0/2)=\hat{U}(T_0/2) \hat{T} \Psi(\bm{r},0).
\eeqn
On the other hand, we have proved in the maintext that 
\beqn\label{S1}
\hat{T}\Psi(\bm{r},0)=\zeta \Psi(\bm{r},T_0/2)=\zeta \hat{U}(T_0/2)\Psi(\bm{r},0), \ \ \zeta=\pm 1.
\eeqn
Multiplying $\hat{U}(T_0/2)$ on both sides of Eq.~\eqref{S1}, we have
\beqn\label{S2}
\hat{U}(T_0/2)\hat{T}\Psi(\bm{r},0)=\zeta \hat{U}^2(T_0/2)\Psi(\bm{r},0).
\eeqn
According to Eq.~\eqref{S0}, the Eq.~\eqref{S2} becomes
\beqn\label{S3}
\hat{T}\Psi(\bm{r},T_0/2)=\zeta \hat{U}^2(T_0/2)\Psi(\bm{r},0).
\eeqn
Multiplying $\zeta$ on both sides of Eq.~\eqref{S3}, we have 
\beqn\label{S4}
\hat{T} \zeta\Psi(\bm{r},T_0/2)=\zeta^2 \hat{U}^2(T_0/2)\Psi(\bm{r},0).
\eeqn
According to the first equal of Eq.~\eqref{S1}, we have
\beqn\label{S5}
\hat{T}^2 \Psi(\bm{r},0) = \hat{U}^2(T_0/2)\Psi(\bm{r},0).
\eeqn

\section{Majorana wave function and  Majorana spin}
Considering  $\text{FI-SC}$ junction proximity on the edge states of 2DTI,  $h(\hat{p})$ and $\bm{m}(\bm{r})\cdot\sigma$ in Eq.\ref{Ham-1} can be reduced to $v_{\rm f}\hat{p}\sigma_z$ and $(m_{\parallel}(\bm r)e^{i\alpha}+m_z(\bm r))\cdot\sigma$ respectively, where $m_z=|\bm m|\sin\theta$. The position vector $\bm{r}$  is expressed as $\bm{r}=x\bm{e_{\rm x}}$ in this one dimension model. The definitions of the magnetic configuration,  chemical potential  and superconducting  pairing potential in real space take
\beqn
\label{sc}
\Delta_{\text{SC}}(x)=\Delta\Theta(x), m_{\parallel/{\rm z}}(x)=m_{\parallel/{\rm z}}\Theta(-x),
\mu(x)=\mu_{\text{SC}}\Theta(x)+\mu_{\text{FI}}\Theta(-x).
\eeqn
The wave function for the electron and hole band in FI region($x<0$) and SC region($x>0$) are  straightforward to show that
\beqn
&&\Psi_{\text{FI}}^{\rm e}(x)=(v_{\rm f}p+m_{{\rm z}}+\mu_{\text{FI}}+E,m_{\parallel}e^{i\alpha},0,0)e^{i{k}_{\text{FI}}^{\rm e}x}, \Psi_{\text{FI}}^{\rm h}(x)=(0,0,-v_{\rm f}p+m_{\rm z}-\mu_{\text{FI}}+E,m_{\parallel}e^{i\alpha})e^{i{k}_{\text{FI}}^{\rm h}x}(x<0),\nonumber\\
&&\Psi_{\text{SC}}^{\rm e}(x)=(\frac{v_{\rm f}p-\mu_{\text{SC}}+E}{\Delta},0,1,0)e^{i{k}_{\text{SC}}^{\rm e}x}, \Psi_{\text{SC}}^{\rm h}(x)=(0,\frac{-v_{\rm f} p-\mu_{\text{SC}}+E}{\Delta},0,1)e^{i{k}_{\text{SC}}^{h}x}(x>0),
\eeqn
 where wave vectors $k_{\text{FI/SC}}^{e/h}$ are defined as
\beqn
 {k}_{\text{FI}}^{\rm {e/h}}=\frac{- i\sqrt{m_{\parallel}^2-(E\pm\mu_{\text{FI}})^2}\mp m_{{\rm z}}}{\hbar v_{\rm f}},{k}_{\text{SC}}^{\rm {e/h}}=\frac{i\sqrt{\Delta^2-E^2}\pm\mu_{\text{SC}}}{\hbar v_{\rm f}}.
\eeqn

Considering the zero energy solution of this BDG Hamiltonian, the wave functions in FI and SC region respectively can be written as
\beqn
&&\Psi_{\text{FI}}(x)=a_{\rm e}(e^{-ik_{\rm m}x},e^{i(\alpha+\varphi)}e^{-ik_{\rm m}x},0,0)+a_{\rm e}^{\ast}(0,0,e^{-i(\alpha+\varphi)}e^{ik_{\rm m}x},-e^{ik_{\rm m}x})^{T}e^{k_{\text{FI}}x}(x<0),\nonumber\\
&&\Psi_{\text{SC}}(x)=m_{\rm e}(ie^{ik_{\text{sc}}x},0,e^{ik_{\text{sc}}x},0)+m_{\rm e}^{\ast}(0,e^{-ik_{\text{sc}}x},0,ie^{-ik_{\text{sc}}x})^{T}e^{-K_{\text{SC}}x}(x>0),
\eeqn
where these parameters $\varphi, k_{\text{FI}}, k_{\rm m}, K_{\text{SC}},k_{\text{sc}} $ are defined for simplification as
\beqn
e^{i\varphi}=\frac{i\sqrt{m_{\parallel}^2-\mu_{\text{FI}}^2}+\mu_{\text{FI}}}{m_{\parallel}}, k_{\text{FI}}=\frac{\sqrt{m_{\parallel}^2-\mu_{\text{FI}}^2}}{\hbar v_{\rm f}}, K_{\text{SC}}=\frac{\Delta}{\hbar v_{\rm f}}, k_{\text{m/sc}}=\frac{m_{\rm z}/\mu_{\text{SC}}}{\hbar v_{\rm f}}.
\eeqn
Coefficients $a_{\rm e}, m_{\rm e}$ are determined by matching the boundary condition $\Psi_{\text{FI}}(0)=\Psi_{\text{SC}}(0)$. The results of this zero energy wave functions take
\beqn
&\Psi_{\text{FI/SC}}(x)=(\psi_{\text{FI/SC}}^{\rm e}(x), i\sigma_y \psi_{\text{FI/SC}}^{\ast\rm e}(x))^{\rm T},\nonumber\\
&\psi_{\text{FI}}^{\rm e}=e^{i(\frac{\pi}{4}-k_{\rm m}x)}(e^{-i\frac{\phi}{2}},e^{i\frac{\phi}{2}}),\psi_{\text{SC}}^{\rm e}=e^{i\frac{\pi}{4}}(e^{-i\frac{\phi}{2}+ik_{\text{sc}}x},e^{i\frac{\phi}{2}-ik_{\text{sc}}x}),
\eeqn
where $\phi=\alpha+\varphi$, the spin of Majorana is calculated by figuring out the Pauli operator average value and takes
\begin{align}
&\langle \mathbf{\sigma}_{FI}^{z}\rangle=0,\langle\mathbf{\sigma}_{\text{FI}}^{x}\rangle=2\cos(\phi)e^{2k_{\text{FI}}x}, \langle \mathbf{\sigma}_{\text{FI}}^{y}\rangle=2\sin(\phi)e^{2k_{\text{FI}}x}(x<0),\\
&\langle \mathbf{\sigma}_{\text{SC}}^{z}\rangle=0,\langle\mathbf{\sigma}_{\text{SC}}^{x}\rangle=2\cos(\phi-2k_{\text{sc}}x)e^{-2K_{\text{SC}}x},
\langle\mathbf{\sigma}_{\text{SC}}^{y}\rangle=2\sin(\phi-2k_{sc}x)e^{-2K_{\text{SC}}x}(x>0).
\end{align}

\section{F and R matrix }
\begin{figure}
\centering
\includegraphics[width=5.0in]{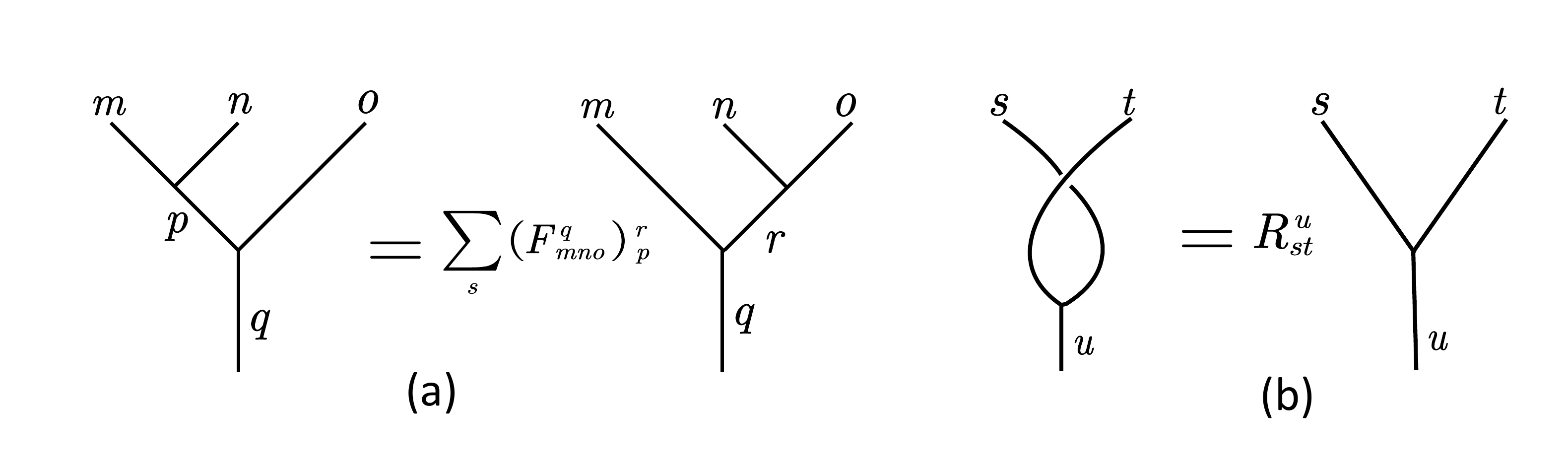}
\caption{(a):The different fusion basis on the left and right is connected by the F-matrix $(F_{mno}^q)_p^r$, where p is the fusion result of anyon m,n and r is the fusion result of n,o. (b):When exchange anyon s and t, R matrix $R_{st}^{u}$ gives different phase with different fusion result. }
\label{FR}
\end{figure}

As Fig.~\ref{MZM-spin} (d) shows in main text, the particles $1,2,3,4$ are MZMs( label as $\sigma$) and  the fusion result of all four MZMs  (label as 5) are set to be vacuum. In the even parity subspace,  matrix multiplication $T=\sum_{\rm c}(F_{\rm{12b}}^{\rm 5})_{\rm a}^{\rm c}R_{\rm {2b}}^{\rm c}(F_{\rm {342}}^{\rm c})_{\rm b}^{\rm d}(F_{\rm {d31}}^{\rm 5})_{\rm c}^{\rm e}$ transforms the  basis $(|00\rangle_{\text{SC}},|11\rangle_{\text{SC}})^{T}$ to the  basis $(|00\rangle_{\text{FI}}, |11\rangle_{\text{FI}})^{T}$,  where F matrix and R matrix provide unitary transformation between different fusion spaces and are defined as the Fig. \ref{FR}~.

 According to the fusion rule of Ising anyon $\sigma \times \sigma=\bm 1+\Psi, \bm 1\times \sigma=\sigma, \Psi\times \sigma=\sigma$\cite{Nayak2008},  both a and b or e and d are same particle required by the total even fermion parity. Particle c has no choice to be anyon $\sigma$.

 F matrixs $(F_{\rm{12b}}^{5})_{\rm a}^{\rm c}, (F_{\rm {d31}}^{\rm 5})_{\rm c}^{\rm e}$ either $(F_{\rm{12b}}^{5})_{\rm a}^{\rm c}=(F_{{\sigma\sigma} \bm{1}}^{\bm 1})_{\bm 1}^{\rm{\sigma}}, (F_{\rm {d31}}^{\rm 5})_{\rm c}^{\rm e}=(F_ {\bm{1}\sigma\sigma}^{\bm 1})_{\rm{\sigma}}^{\bm 1} $ with  $a=b=d=e=\bm 1$ or  $(F_{\rm{12b}}^{\rm 5})_{\rm a}^{\rm c}=(F_{\rm{\sigma\sigma \psi}}^{\bm 1})_{\rm{\psi}}^{\rm{\sigma}},(F_{\rm{d31}}^{\rm 5})_{\rm c}^{\rm e}=(F_{\rm{\psi\sigma\sigma}}^{\bm 1})_{\rm{\sigma}}^{\rm{\psi}}$ with $a=b=d=e=\Psi$. Both of the two cases, $(F_{\rm{12b}}^{5})_{\rm a}^{\rm c}$ and $ (F_{\rm {d31}}^{\rm 5})_{\rm c}^{\rm e}$ are unit matrix. Thus, we have $(F_{\rm{12b}}^{\rm 5})_{\rm a}^{\rm c}= (F_{\rm{d31}}^{\rm 5})_{\rm c}^{\rm e}=\text{I}_{\rm{2\times 2}}$. $R_{\rm{2b}}^{\rm{c}}$ is diagonal matrix and takes\cite{Pachos2012}
\beqn
& R_{\rm{2b}}^{\rm c}=\begin{pmatrix}
  R_{\rm{\sigma 1}}^{\rm{\sigma}} & 0\\
   0 &R_{\rm{\sigma \Psi}}^{\rm{\sigma}}\end{pmatrix}=\begin{pmatrix}
 1 & 0\\
   0& i\end{pmatrix}.
\eeqn
The matrix $(F_{\rm{342}}^{\rm c})_{\rm b}^{\rm d}=(F_{\rm{\sigma\sigma\sigma}}^{\rm{\sigma}})_{\rm b}^{\rm d}$ is standard F matrix of Ising anyon and takes
\beqn
 (F_{\rm{\sigma\sigma\sigma}}^{\rm{\sigma}})_{b}^{d}=\begin{pmatrix}
(F_{\sigma\sigma\sigma}^{\rm{\sigma}})_{1}^{1}&(F_{\rm{\sigma\sigma\sigma}}^{\rm{\sigma}})_{1}^{\rm{\psi}}\\
  (F_{\rm{\sigma\sigma\sigma}}^{\rm{\sigma}})_{\rm{\psi}}^{1}&(F_{\rm{\sigma\sigma\sigma}}^{\rm{\sigma}})_{\rm{\psi}}^{\psi}\end{pmatrix}=\frac{1}{\sqrt{2}}\begin{pmatrix}
1& 1\\
   1& -1\end{pmatrix}.
\eeqn
Thus the transformation matrix is
\beqn
T&&=\sum_{\rm c}(F_{\rm{12b}}^{\rm 5})_{\rm a}^{\rm c}R_{\rm {2b}}^{\rm c}(F_{\rm {342}}^{\rm c})_{\rm b}^{\rm d}(F_{\rm {d31}}^{\rm 5})_{\rm c}^{\rm e}\nonumber\\
&&=\text{I}_{\rm{2\times 2}}R_{\rm{2b}}^{\rm c}(F_{\rm{342}}^{\rm c})_{\rm b}^{\rm d}\text{I}_{\rm{2\times 2}}\nonumber\\
&&=\frac{1}{\sqrt{2}}\begin{pmatrix}
  1 & 0\\
   0 & i\end{pmatrix} \begin{pmatrix}
1& 1\\
   1& -1\end{pmatrix}\nonumber\\
&&=\frac{1}{\sqrt{2}}\begin{pmatrix}
  1 & 1\\
   i& -i\end{pmatrix}.
\eeqn

\section*{Numerical Results}
We adopt the Bernevig-Hughes-Zhang (BHZ) model, to show the robustness of the braiding process against various kinds of disorder effects and the respective success rate when non-adiabatic or finite size effect are brought into play.
\subsubsection{BHZ model}
The Hamiltonian of the BHZ model can be written as
\begin{equation}
\mathcal{H}_{\text{2DTI}}\left(\vec{k}\right)=\left( \begin{array}{cc}
\vec{d}(\vec{k})\cdot\vec{\sigma}& 0\\
0& \vec{d}^*(-\vec{k})\cdot\vec{\sigma}
\end{array} \right)
\end{equation},
where $\sigma$'s are the Pauli matrices operating on orbitals. The top-left block describes state with spin-up and the bottom-right block describes its time-reversal partner, and $\vec{d}(\vec{k})$ is given by
\begin{eqnarray}
\vec{d}(\vec{k})&=&\left(2A\sin(k_x), 2A\sin(k_y), \Delta-4B \sin^2(k_x)-4B\sin^2(k_y) \right) \\
&=&\left(2A\sin(k_x), 2A\sin(k_y), (\Delta-4B) +2B \cos(k_x)+2B\cos(k_y) \right)
\end{eqnarray}

The Hamiltonian satisfies time reversal symmetry $\Theta = i s_y K$, where $K$ is the complex conjugate operator. In order to obtain the helical edge states, we set open boundary in y direction while keeping the periodic boundary condition in x direction, that is,

\begin{eqnarray}
H_{\text{2DTI}}\left(k_x\right)&=&\sum_y  C_{k_x,y}^\dagger \left(\left((\Delta-4B) +2B \cos(k_x)\right) \sigma_z \otimes s_0 + 2A\sin(k_x) \sigma_x \otimes s_z) C_{k_x,y}^\dagger \right) \nonumber\\
&&+\left[ C_{k_x,y}^\dagger \left( B\sigma_z\otimes s_0  C_{k_x,y+1} -iA \sigma_y \otimes s_0  C_{k_x,y+1} + iA \sigma_y \otimes s_0  C_{k_x,y-1}\right) + h.c.\right],
\end{eqnarray}
where $C_{k_x,y}^\dagger =\left( c_{k_x,y,A\uparrow}^\dagger, c_{k_x,y,A\downarrow}^\dagger, c_{k_x,y,B\uparrow}^\dagger, c_{k_x,y,B\downarrow}^\dagger \right)$ and $s$'s are the Pauli matrices operating on spins. \\

Analytically, the helical edge states have energy $\pm 2Ak_x$ and have the form
\begin{eqnarray}
\psi_{\text{edge}} &=& \mathcal{A}_{\pm} \left(e^{\kappa_+ x} - e^{\kappa_- x}\right) u_{\pm} \left(k_y\right); \nonumber\\
u_+ &=& \left( \begin{matrix}
1 \\ 1
\end{matrix} \right)_\sigma \otimes
\left( \begin{matrix}
1 \\ 0
\end{matrix} \right)_s; \nonumber\\
u_- &=& \left( \begin{matrix}
1 \\ 1
\end{matrix} \right)_\sigma \otimes
\left( \begin{matrix}
0 \\ 1
\end{matrix} \right)_s,
\end{eqnarray}
where $\kappa_{\pm} = i/B \times \sqrt{2A^2+B^2 k_x^2 + B \Delta \pm 2A\sqrt{A^2+B^2 k_x^2 - B\Delta}}$ and the subscripts $\sigma$ and $s$ denote the subspace of sublattice and of spin respectively. In particular, when $k_x=0$ and $\Delta/A$ is small, the edge mode decays at a rate approximately equal to $e^{\Delta x/(2A)}$.\\

From the expression of the edge mode, it can be seen that a magnetic field acting in the direction orthogonal to $z$-axis gaps out the helical edge modes into trivial ferromagnetic insulators and a pairing gaps out the metallic edge modes into a 1D topological superconductor akin to Kitaev's model. Moreover, at the boundaries of the induced topological superconducting and trivial ferromagenetic insulator arise Majorana bound states.

\subsubsection{Numerical time-evolution}
The time evolution of the Majorana wave functions in the first quantization language can be tracked by explicitly solving the time-dependent BdG Schordinger Equation:
\begin{equation}\label{t-dBdG}
i\hbar\frac{\partial}{\partial t}\psi(t)=\mathcal{H}_{\text{BdG}}(t)\psi(t),
\end{equation}
 where $\psi(t)$ is the wavefunction of the Majorana operator in Nambu basis $( \psi_i, \psi_i^\dagger)^T$. The solution to~(\ref{t-dBdG}) is given by
\begin{equation}
\psi(t)=\mathcal{T} \exp[-\frac{i}{\hbar}\int_0^t dt\mathcal{H}(t)]\psi(0),
\end{equation}
where $\mathcal{T}$ is the time-ordering operator.\\

In numerically simulation, we can carry out the time evolution step by step in order to get rid of the time-ordering operator which is hard to handle:
\begin{equation}\label{evolstep}
\psi(t) = \lim_{N\rightarrow \infty}\prod_{n=0}^{N-1} \exp\left[-\frac{i}{\hbar} \frac{t}{N}\mathcal{H}\left(\frac{nt}{N}\right)\right]\psi(0).
\end{equation}

\subsubsection{Evolution for a full braiding}
In this part, the robustness of the full braiding is verified by introducing static disorder, dynamical out-of-plane magnetization fluctuation. And in order to investigate the experimental feasibility of our proposal, we consider the error that may be caused by finite size effect and non-adiabatic braiding, showing that the outcome of our set up is rather reliable in a wide range of parameter regime.
In our numerical study, the Hamiltonian can be written as
\begin{eqnarray}
H_{\text{2DTI}}&=&\sum_{xy}  C_{x,y}^\dagger (\Delta-4B)\sigma_z C_{x,y} +\left[ C_{x,y}^\dagger \left( B \sigma_z \left( C_{x+1,y}+ C_{x,y+1} \right) + A \sigma_y  C_{x+1,y}  + A \sigma_x \otimes s_z  C_{x,y+1} \right) + h.c.  \right] \\
&&+ i \alpha_{soc}\sum_{xy}  C_{x,y}^\dagger s_x  (\frac{3}{4} C_{x+1,y}+ \frac{1}{4} C_{x+3,y})+h.c.\\
&& + \sum_{\{x,y\} \in M} C_{x,y}^\dagger \vec{m} \cdot \vec{s} C_{x,y} + \sum_{\{x,y\} \in S} \left[ C_{x,y}^\dagger \Delta_s i s_y C_{x,y}^\dagger + h.c. \right]
\end{eqnarray},
where $C_{x,y}^\dagger =\left( c_{x,y,A\uparrow}^\dagger, c_{x,y,A\downarrow}^\dagger, c_{x,y,B\uparrow}^\dagger, c_{x,y,B\downarrow}^\dagger \right)$, and $M$ and $S$ are the FI and TSC region respectively. Here, we take $A=2, \Delta = 1.6, B =1,  m = 0.8, \Delta_s = 0.3, \mu=0.4, \alpha_{soc}=0.25$. The $\alpha_{soc}$ term denotes the spin-orbit coupling proportional to $\sin^3(k_x)$. The magnetization and superconducting pairing  are induced on the edge of the system with the depth of $3$ lattice sites.

The intrinsic spin of MF is determined by the direction of the FI magnetization. And by rotating the magnetization adiabatically, the spin of the MF is rotated correspondingly, leading to an effective braiding operation on the MFs. Let $\vec{m}(t)=(m \cos(\omega t) , m \sin(\omega t), 0)$, where $\omega=2\pi/T_0$. The magnetization direction is rotated by $2\pi$ during time $T_0$. After rotating $\vec{m}$ by  $\vec{m}\to -\vec{m} \to \vec{m}$ without closing the gap, $\gamma_2 \to -\gamma_2, \gamma_3 \to -\gamma_3$, that is, $\langle\gamma_2(T_0)|\gamma_2(0)\rangle=-1$, as shown in Fig.~\ref{braiding}(b)(black, solid). In order to verify the robustness of the MF braiding, we take into account the static disorder potential and dynamical out-of-plane magnetization fluctuation. The disorder potential $V_{dis}$ is  given by $V_{dis}=\sum_{i} W_{i} n_{i} $, where $n_i=\sum_{s,\sigma} n_{i,s,\sigma}$ and $W_i$ distributes randomly within the range of $[-W,W]$. As can be shown in Fig.~\ref{braiding}(b)(red, dotted), the non-Abelian braiding is immune to static disorder. In a realistic experimental braiding operation, magnetization direction may tilt away from $x-y$ plane during the braiding process, resulting in a time dependent magnetization $\vec{m}(t)=|\bm{m}|\bigr[\cos\delta(t) \cos(\omega t) ,  \cos\delta(t) \sin(\omega t),  \sin\delta(t)\bigr]$. When a dynamical tilting $\delta(t)=\frac{\pi}{4}\sin^2(\omega t)$ is induced, variation may be induced during the braiding process, yet the final result of non-Abelian braiding $\langle\gamma_2(T_0)|\gamma_2(0)\rangle=-1$ remains unchanged(blue,dashdotted). Other $\delta(t)$ configurations have also been tested, including linear variation and adiabatic random $\delta(t)$. Viariations in the middle of the braiding operation vary for different $\delta(t)$ configurations, yet for all configurations $\langle\gamma_2(T_0)|\gamma_2(0)\rangle=-1$ remains unchanged. In fact, so long as the tilting of $\delta$ does not close the gap, the non-Abelian braiding is maintained during an adiabatic braiding operation and the outcome of a braiding operation is independent of the path of the magnetization during the process. \\

We now proceed to study the physical effects that may cause error in braiding operation. In the adiabatic limit, the system transforms inside the degenerate ground state subspace and braiding is topologically protected. However, if the braiding operation is carried out too quickly, error may be induced in braiding.
In Fig.~\ref{braiding}.(c), fidelity $F$ versus the length of FI region $L_{FI}$ relation is given for $T_0 \Delta_{SC}= 10, 30, 100 $ (blue, red, green), where $\Delta_{SC}$ denotes the superconducting gap. The inset shows that for error less than $2\%$, the braiding time must be longer than $30$ times the bulk gap $\Delta_{SC}$. Take the typical braiding time $ T_0\Delta_{SC}= 30$ as an example. In order to make sure that the fidelity is no less than $99\%$, the distance between the 2 MZM modes must be further than $L_{FI}\approx 6 \xi$, where $\xi$ is the Majorana wavefunction localization length in the FI region. As we can see from the wavefunction above, the localization length of MZM inside the FI region is gicen by $\xi_=\frac{\hbar v_f}{\sqrt{|m|^2-\mu^2}}$. In the HgTe quantum well, the fermion velocity of edge states is $\hbar v_f=0.36\textrm{meV} \cdot \mu \textrm{m}$. For $|m|\approx 3 \textrm{meV} $ and dopping $\mu=1\textrm{meV} $, the length of the FI should be no shorter than $0.8 \mu \textrm{m}$.

\section{charge pumping influenced by temperature}
Considering the length of FI  is finite in FI-SC junction and label as L, the  edge states on the left of FI region can be viewed as metal lead. We label this region as QSH region($x<-L$).
When rotating the magnetization of FI adiabatically, the charge pumping takes \cite{Avron2004}
\beqn
\Delta Q=-\frac{i}{2\pi}\int_{0}^{T}dt\int_{-\infty}^{\infty} dE\rho^{'}(E)\Xi_{\rm {jj}}(E,t),
\eeqn
where $\Xi(E,t)=i\frac{dS(E,t)}{dt}S^\ast(E,t)$ is the energy shift, $\rho^{'}(E)$ is the derivative of Fermion-Dirac function over energy. In our model of QSH-FI-SC hybrid system,  the scattering matrix $S(E,t)$ takes
\beqn
   S(E,t)=&\begin{pmatrix}
r_{\rm {ee}}(E,t)&r_{\rm {eh}}(E,t)\\
r_{\rm {he}}(E,t)&r_{\rm {hh}}(E,t)\end{pmatrix},
\eeqn
where $r_{\rm {ee}}$ is normal reflection coefficient and $r_{\rm {eh}}$ is Andreev reflection coefficient. The scattering coefficients $r_{\rm{hh}},r_{\rm{he}}$ for incident hole from QSH region are connected to the coefficients $r_{\rm {ee}},r_{\rm {eh}}$ for incident electron  by particle-hole-symmetry. The result shows $r_{\rm {ee}}(E,t)=r_{\rm {hh}}^{\ast}(-E,t),r_{\rm {eh}}(E,t)=r_{\rm {he}}^{\ast}(-E,t)$. Thus the charge pumping for electron or hole takes
\beqn\label{delta-q}
&&\Delta Q_{\rm{e}}=-\frac{i}{2\pi}\int_{0}^{T}dt\int_{-\infty}^{\infty}dE\rho^{'}(E)(\frac{\partial r_{\rm{ee}}}{\partial t}r_{\rm{ee}}^{\ast}+\frac{\partial r_{\rm{eh}}}{\partial t}r_{\rm{he}}^{\ast}),\nonumber\\
&&\Delta Q_{\rm{h}}=-\frac{i}{2\pi}\int_{0}^{T}dt\int_{-\infty}^{\infty}dE\rho^{'}(E)(\frac{\partial r_{\rm{hh}}}{\partial t}r_{\rm{hh}}^{\ast}+\frac{\partial r_{\rm{he}}}{\partial t}r_{\rm{eh}}^{\ast}).
\eeqn

 Now, we  calculate the scattering coefficients $r_{\rm{ee}}, r_{\rm{eh}}$ as follows.  The definitions of the magnetic configuration, chemical formula and superconducting  pairing potential in (\ref{sc}) are  redefined as
\beqn
&\Delta_{\text{SC}}(x)=\Delta\Theta(x), m_{\parallel/z}(x)=m_{\parallel/z}\Theta(-x)\Theta(x+L) ,\nonumber\\
&\mu(\bm{r})=\mu_{\text{SC}}\Theta(x)+\mu_{\text{FI}}\Theta(-x)\Theta(x+L)+\mu_{\text{QSH}}\Theta(-x-L).
\eeqn
Without loss of generality,  chemical potential in FI region is tuned as $\mu_{\text{FI}}=0$ for simplification.   The scattering states in QSH, FI and SC regions  take
\beqn
&\Psi_{\text{QSH}}(x)=(1,0,0,0)e^{ik_1x}+r_{ee}(0,1,0,0)e^{-ik_1x}+r_{eh}(0,0,1,0)e^{ik_2x}(x<-L),\nonumber\\
&\Psi_{\text{FI}}(x)=a(e^{i\theta_m}e^{-i\alpha},1,0,0)e^{-k_{\text{FI}}x-ik_{\rm m}x}+b(e^{-i\theta_m}e^{-i\alpha},1,0,0)e^{k_{\text{FI}}x-ik_{\rm m}x}\nonumber\\
&+c(0,0,e^{i\theta_m}e^{-i\alpha},1)e^{k_{\text{FI}}x+ik_{\rm m}x}+d(0,0,e^{-i\theta_m}e^{-i\alpha},1)e^{-k_{\text{FI}}x+ik_{\rm m}x}(-L<x<0),\nonumber\\
&\Psi_{\text{SC}}(x)=m(e^{i\beta},0,1,0)e^{ik_{\text{sc}}x-K_{\text{SC}}x}+n(0,e^{-i\beta},0,1)e^{-ik_{\text{sc}}x-K_{\text{SC}}x}(x>0).
\eeqn
Above parameters $k_{1}, k_{2},\theta_m,\beta$in the wave functions are defined as
\beqn
k_{1}=\frac{E+\mu_{\text{QSH}}}{\hbar v_{\rm f}}, k_{2}=\frac{\mu_{\text{QSH}}-E}{\hbar v_{\rm f}},\theta_m=\arccos{\frac{E}{m_{\parallel}}},\beta=\arccos{\frac{E}{\Delta}}.
\eeqn
Analytically, normal reflection coefficient and  Andeev reflection coefficient can be obtained by matching boundary conditions $\Psi_{\text{QSH}}(-L)=\Psi_{\text{FI}}(-L),\Psi_{\text{FI}}(0)=\Psi_{\text{SC}}(0)$.
\beqn
\label{r}
&&r_{ee}=\frac{e^{i(\alpha+\theta_m)}(-1+e^{2k_{\text{FI}}L})(-e^{2i\beta}+e^{2i\theta_m}-e^{2k_{\text{FI}}L}+e^{2i\beta+2i\theta_m+2k_{\text{FI}}L})}
{e^{2i\beta}-e^{2i\theta_m}+2e^{2i\theta_m+2k_{\text{FI}}L}-2e^{2i\beta+2i\theta_m+2k_{\text{FI}}L}-e^{2i\theta_m+4k_{\text{FI}}L}+e^{2i\beta+4i\theta_m+4k_{\text{FI}}L}},\nonumber\\
&&r_{eh}=\frac{e^{-2ik_{\rm m}L+i\beta+i(k_2-k_1)L+2k_{\text{FI}}L}(-1+e^{2i\theta_m})^2}{e^{2i\beta}-e^{2i\theta_m}+2e^{2i\theta_m+2k_{\text{FI}}L}-2e^{2i\beta+2i\theta_m+2k_{\text{FI}}L}-e^{2i\theta_m+4k_{\text{FI}}L}+e^{2i\beta+4i\theta_m+4k_{\text{FI}}L}}.
\eeqn

For integration calculation in Eq.(\ref{delta-q}),  specific trajectory have to be assigned.   The trajectory that magnetization rotate around $\bm{e_z}$ axis with $\theta=\pi/2$ is chosen  in our calculation. In this evolution path, only  azimuth angle $\alpha$ is time-dependent.  Consequently, the derivative $\frac{\partial r_{\rm{eh/he}}}{\partial t}$ is zero and electron or hole pumping when rotating magnetization $2\pi$ shows
\beqn\label{q}
\Delta Q_{\rm {e/h}}&&=\pm\frac{1}{2\pi}\int_{0}^{T}\frac{\partial\alpha}{\partial t} dt \int_{-\infty}^{\infty}\rho^{'}(E)r_{\rm {ee/hh}}^2dE\nonumber\\
&&=\pm \frac{(\alpha(T)-\alpha(0))}{2\pi} \int_{-\infty}^{\infty}\rho^{'}(E)r_{\rm {ee/hh}}^2dE\nonumber\\
&&=\pm \int_{-\Delta}^{\Delta}\rho^{'}(E)r_{\rm {ee/hh}}^2dE.
\eeqn
The upper (lower) sign refers to electron (hole) pumping. The whole charge pump into the superconductor takes the form
\beqn
\Delta Q^{'}&&=\frac{\Delta Q_{\rm {e}}-\Delta Q_{\rm {h}}}{2}\nonumber\\
&&=e\int_{-\Delta}^{\Delta}\rho^{'}(E)r_{\rm{ee}}^2dE.
\eeqn
It is noted that in our calculation, the  range for energy  integration is seted in the superconductor gap in order to guarantee that the charge  pump into the nonlocal fermion level in topological superconductor.


\begin{thebibliography}{56}
\bibitem [{top()}]{topspin}%
  \BibitemOpen
  \href@noop {} {}\bibinfo {note} {It should be noted that the word "spin" in
  spin-statistics theorem is refereed to the topological spin of anyons and
  should be distinguished from the actual spin. For example the topological
  spin of MZM is 1/8 but the Majorana wave function is a spinor which only
  takes half integer value.}\BibitemShut {Stop}%
\end{thebibliography}
\end{document}